\documentclass[floatfix,aip,amsmath,amssymb,reprint]{revtex4-1}
\usepackage{graphicx}
\usepackage{amssymb}
\usepackage{color}
\usepackage{epsfig}
\usepackage{amsmath}
\usepackage{placeins}
\usepackage[caption=false]{subfig}
\usepackage{lmodern}  
\usepackage{lipsum}
\usepackage{balance}
\usepackage[T1]{fontenc}
\usepackage{todonotes}
\usepackage{ulem}

\usepackage[utf8]{inputenc}
\usepackage{amssymb}
\DeclareUnicodeCharacter{211D}{\mathbb{R}}

\begin{document}

\preprint{AIP/123-QED}

\title{Fragility in Glassy Liquids: A Structural Approach Based on Machine Learning}
\author{Indrajit Tah}
\email{itah@sas.upenn.edu}

\author{Sean A. Ridout}%
 \email{ridout@sas.upenn.edu }

\author{Andrea J. Liu}
 \email{ajliu@upenn.edu}
 
\affiliation{$^1$ Department of Physics and Astronomy,
University of Pennsylvania 209 South 33rd Street
Philadelphia, PA, USA }

\date{\today}

\begin{abstract}
The rapid rise of viscosity or relaxation time upon supercooling is universal hallmark of glassy liquids. The temperature dependence of the viscosity, however, is quite non universal for glassy liquids and is characterized by the system's ``fragility,'' with liquids with nearly Arrhenius temperature-dependent relaxation times referred to as strong liquids and those with super-Arrhenius behavior referred to as fragile liquids. What makes some liquids strong and others fragile is still not well understood. Here we explore this question in a family of glassy liquids that range from extremely strong to extremely fragile, using ``softness,'' a structural order parameter identified by machine learning to be highly correlated with dynamical rearrangements. We use a support vector machine to identify softness as the same linear combination of structural quantities across the entire family of liquids studied. We then use softness to identify the factors controlling fragility.
\end{abstract}

\maketitle
\section{\bf Introduction}
\label{section:1}

The extremely rapid increase in viscosity or relaxation time when a liquid cools is a universal hallmark of glassy liquids \cite{RevModPhys.83.587,doi:10.1146/annurev-conmatphys-031113-133848,doi:10.1126/science.1120714,Karmakar_2015,acsomega.0c04831}. The temperature dependence of relaxation time, which is characterized by its ``fragility,'' \cite{Angell1924} is, however, non-universal, differing widely among glassy liquids. Glassy liquids which show an Arrhenius temperature dependence (exponential in inverse temperature) in relaxation time are called strong glassformers and those with stronger-than-Arrhenius (super-Arrhenius) dependence on temperature are known as fragile glassformers. An infamous property of the glass transition is that the increase of relaxation time upon cooling is unaccompanied by any qualitative change in the liquid structure, and very little quantitative change in it. As a result, structural insight into the differences between strong and fragile liquids has been lacking.

Many experimental \cite{Mauronatcomm,wei_linking_2015,doi:10.1063/1.4964362} and numerical \cite{sun_structural_2022,sastry_nature,Kaori_nature,PhysRevLett.126.208001} studies have been conducted to better understand the correlation of fragility to structure. 
Recent efforts to characterize structure in terms of locally preferred structures \cite{doi:10.1063/1.2773716,Hallett,PhysRevX.8.011041}, medium-range crystalline order \cite{PhysRevLett.99.215701,tanaka_2010,Kawasaki_2011,PhysRevLett.121.085703}, and complex topological clusters~\cite{Mauro201131} have been successful for some systems but have not been applied across the strong-to-fragile spectrum of glassformers. However, a machine learning approach \cite{cubuk_identifying_2015,Schoenholz_Nat_Physics} shows promise. This approach identifies a microscopic structural variable called ``softness'' \cite{Schoenholz_Nat_Physics} that describes each particle's local structural environment in a way that is designed to be strongly correlated with its dynamics. It has been applied successfully across a spectrum of systems \cite{cubuk_identifying_2015,Schoenholz_Nat_Physics,Schoenholz263,doi:10.1126/science.aai8830,Sharp10943,ma_heterogeneous_2019,harrington_machine_2019,PhysRevE.101.010602,cubuk_unifying_2020,D0SM01575J,PhysRevX.11.041019,https://doi.org/10.48550/arxiv.2011.13049}.
Most important for our purpose here, softness appears to be associated with the typical energy barrier for rearrangement, with particles of higher softness having lower barriers. This association between structure and typical energy barrier holds across systems spanning a broad range of disorder spanning from glassforming liquids~\cite{Schoenholz_Nat_Physics} to crystals~\cite{Sharp10943} as well as a wide range of fragilities ranging from fragile (super-Arrhenius) behavior~\cite{Schoenholz_Nat_Physics} to strong (Arrhenius)~\cite{cubuk_unifying_2020} and even beyond (sub-Arrhenius behavior)~\cite{D0SM01575J}. Thus the softness of a particle appears to provide a useful structural measure of the typical energy barrier that must be surmounted for the particle to rearrange.

Here we extend the interpretation that softness provides a measure of such energy barriers to study the implied entropic barriers as well for a family of systems that span a wide range of fragilities. In order to study behavior across the family, we use the same definition of softness for all the systems. 
We show that the average softness $\langle S \rangle$ and free energy barrier $\Delta F(\langle S \rangle)$ are stronger functions of temperature for fragile liquids than strong ones. Moreover, the deviation of the relaxation time from simple proportionality to $1/P_R(\langle S\rangle)$, where $P_R(S)$ is the probability that a particle of softness $S$ will rearrange, is compressed into a smaller range of temperatures relative to $T_g$ as fragility increases. Together, these three factors lead to a much stronger increase of the relaxation time with decreasing temperature for fragile liquids than for strong ones. 

Sec.~\ref{section:2} describes the model and simulation details, Sec.~\ref{section:3} describes the machine learning analysis methodology, Sec.~\ref{section:4} contains our results and Section~\ref{section:5} summarizes our conclusions. 

\section{\bf Model and Simulations}
\label{section:2}
We have conducted equilibrium molecular dynamics (MD) simulations in three dimensions for model glass forming liquids with harmonic repulsive interactions. Specifically, we study $50:50$ binary mixtures of $N = 10976$ particles that interact via the harmonic pair potential \cite{Durian_PRL1995,PhysRevLett.88.075507,doi:10.1146/annurev-conmatphys-070909-104045,Berthier_2009}: 
\begin{equation}
V_{\alpha\beta}(r) = \epsilon_{\alpha\beta} \left ( 1-\frac{r}{\sigma_{\alpha\beta}}\right)^2
\end{equation}
for $r < \sigma_{\alpha\beta}$ and $V_{\alpha\beta}(r) = 0$ otherwise. 
Here $(\sigma_{\alpha\alpha}, \sigma_{\alpha\beta}, \sigma_{\beta\beta})$ = $(1.00,\ 1.20,\ 1.40)$, and   ($\epsilon_{\alpha\alpha}, \epsilon_{\alpha\beta}, \epsilon_{\beta\beta}) = (1.0,\ 1.0,\ 1.0) .$ Each simulation is run at fixed number density ($\rho$) and temperature ($T$).  All data was collected after the system reaches thermal equilibrium. This model shows a strong increase of fragility with increasing density \cite{Berthier_2009,kinetic_fra}.  In particular, at low densities near the jamming transition the relaxation time exhibits Arrhenious temperature dependence (strong behavior) while at high density the relaxation time is super-Arrhenious (fragile behavior)~\cite{Berthier_2009}. We performed our simulations in the  density range $\rho \in [0.65, 0.82]$ to span a wide range of fragilities. In the following, length is measured in units of $\sigma_{\alpha\alpha}$, energy in units of $\epsilon_{\alpha\alpha}$, temperature in units of $\epsilon/\kappa_B$, time scale in units of $(\frac{m\sigma_{\alpha\alpha}^2}{\epsilon_{\alpha\alpha}})^{1/2}$,  masses and $\kappa_B$ are set to unity.

\section{\bf Softness Analysis}
\label{section:3}
Following earlier work, we use machine learning to identify a structural variable, softness, that characterizes the local structural environment surrounding each particle and correlates strongly with rearrangements. To do this, we must first characterize the local structural environment, then identify rearranging and non-rearranging particles for the training set. We then use a Support Vector Machine (SVM) \cite{SVM1,SVM2} to construct a hyperplane that best separates the rearranging particles in the training set from the non-rearranging ones.

In contrast to previous work~\cite{Schoenholz_Nat_Physics,Schoenholz263,Sussman10601} we do not apply the softness analysis to inherent structures (configurations obtained by quenching snapshots to zero temperature). Instead, we apply it directly to configurations of the thermal system. We find that this diminishes the classification accuracy slightly (to around $82\%$) compared to typical results for inherent structures in Lennard-Jones systems (around 88-90\%) but that softness still captures enough to be useful.

The local structural environment is quantified by three sets of structure functions. The first and second sets of structure functions are  closely related and are inspired by the symmetry functions that were proposed by Behler and Parrinello \cite{PhysRevLett.98.146401}. Both these structural descriptors are widely used  to quantify the particle local structure environment in previous studies \cite{cubuk_identifying_2015,Schoenholz_Nat_Physics,doi:10.1126/science.aai8830,Schoenholz263,Sussman10601,Sharp10943,PhysRevX.11.041019}. The first set of structural function measures the local density of a reference particle $\kappa$ at distance $\mu \pm L$.
\begin{equation}
G(\kappa;\mu) = \sum_i \exp^{-(r_{ik}-\mu)^2/L^2}
\label{density}
\end{equation}
$\kappa$ is the reference particle whose local environment we want to probe. 
$\mu$ is the distance at which we want to measure the local density of particle $\kappa$. $r_{ik}$ is the distance between particle $\kappa$ and $i$. $G(\kappa;\mu)$ provides the radial density  of reference particle $\kappa$ at distance $\mu$. Here we keep $L = 0.1 \sigma_{\alpha\alpha}$ fixed. We take $\mu$ in between 
$0$ to $3.0 \sigma_{\alpha\alpha}$ with an increment of $0.1 \sigma_{\alpha\alpha}$. The second set of structural descriptor captures bond angle information.
\begin{equation}
\Psi(\kappa;\xi,\lambda,\zeta)=\sum_{i,j} \exp^{-(r_{i\kappa}^2+r_{j\kappa}^2+r_{ij}^2)/\xi^2} (1+\lambda\cos{\theta_{\kappa ij}})^\zeta
\label{angular}
\end{equation}
where $\theta_{\kappa ij}$ is the angle between particles $\kappa$, $i$ and $j$ and $\lambda = \pm 1$. The angular resolution of the structure functions is controlled by the parameter $\zeta$.  The third set of structural descriptors includes the number of Delauney neighbors for the reference particle $\kappa$, the number of Delauney neighbors that are in contact (within interaction range) and the number of neighbors that are not in contact (beyond the interaction range) with the reference particle $\kappa$~\cite{doi:10.1063/5.0035395,https://doi.org/10.48550/arxiv.2011.13049}. Altogether we retain 73 structural descriptors. Note that this is overkill; our goal is not to obtain the most parsimonious description of local structure that yields useful information, but to have a definition expansive enough to yield high prediction accuracy across the entire range of system densities studied.

Once each particle's structural descriptors ($\mathbf{F_i}$ = $\{F_i^1,..,F_i^M\}$, here M = 73) have been chosen, the next step is to choose the training set, which consists of two equal-sized subsets, one containing rearranging particles and the other subset containing non-rearranging particles. Stochasticity plays a role in whether a particle will rearrange or not; to reduce the effects of stochasticity on the training set, we choose particles from the extremes of the distribution: to be labeled as a rearranging particle it must rearrange within the very short  cage-breaking time scale, $\tau_\beta$ \cite{doi:10.1063/1.5033555} in the future; to be a non-rearranging particle it must not have rearranged for a long time, $10\tau_\alpha$, in the past. We then use support-vector machine (SVM) method  with a linear kernel to construct the hyperplane $w \cdot \mathbf{F} - b =0 $ that best separates the rearranging from the non-rearranging particles. 

The first step is to define what we mean by a rearrangement. We calculate $D_{min}^2(i)$ \cite{PhysRevE.57.7192}, the non-affine displacement of a particle $i$
\begin{multline}
D^2_{\mathrm{min}}(i,t,\Delta t )= \underset{ \epsilon_{\alpha \beta}}{\text{min}}\Big\{ \sum_\alpha \sum_j \big[ (r^\alpha_j(t)-r^\alpha_i(t))-\\ \sum_{\beta}(\delta_{\alpha \beta} -\epsilon_{\alpha \beta}) (r^\beta_j(0)-r^\beta_i(0)) \big ]^2 \Big\}
\end{multline}
Here $\alpha$ and $\beta$ denote spatial coordinates and $j$ indexes the neighbors of particle $i$ out to a distance of $R_D = 2.5$.  The first term represents the actual displacements of nearby particles relative to particle $i$, while the second term represents the relative displacements of neighboring particles if they were in a region of uniform strain. We then find the $\epsilon_{\alpha \beta}$ that minimizes $D^2 (t,\Delta t )$. The local deviation from affine deformation during the time interval $t$ and $\Delta t$, referred to as $D_{min}^2$, measures the non-affine component of local deformation. The  time interval $\Delta t$ is set to be longer than the ballistic time scale and close to the cage breaking time scale; we use $\Delta t = 12$. 

We can now define a rearrangement. A particle is rearranging if $D_{min}^2$ is above than a threshold value $D^2_{\mathrm{min},0}$ \cite{cubuk_identifying_2015}.  
We choose $D^2_{\mathrm{min},0}$ so that roughly 1 \% of small particles undergo a rearrangement at its low temperature state point (the low-temperature state point for each density has been chosen so that the systems' relaxation times are relatively similar). Following that, we use the same threshold for all subsequent temperatures for a specific density system. Table ~\ref{table:1} lists the values of $D^2_{\mathrm{min},0}$ for each density.

\begin{table}
\caption{}
 \begin{center}
  \begin{tabular}{ | c | c | c | }
  
 \hline
 Density & $D^2_{\mathrm{min},0}$ \\
 \hline
 $\rho = 0.82$  & 0.28     \\
\hline
$\rho = 0.78$  &  0.26    \\
\hline
$\rho = 0.72$  &  0.24    \\
\hline
$\rho = 0.69$  &  0.235    \\
\hline
$\rho = 0.65$  &  0.21    \\
\hline
\end{tabular}
\label{table:1}
\end{center}
\end{table}

\vskip +0.05in

We show the results for small particles in our binary mixture here, but our results have been verified for large particles (see Appendix ~\ref{Appendix_sce_1}). 

Now that we have defined rearrangements, we must next assemble a training set and construct an SVM hyperplane. Our study requires obtaining a hyperplane that yields high prediction accuracy across the range of temperatures and densities of interest. To do this, we start by assembling a training set from the highest density system studied ($\rho=0.82$) at a low temperature (T = 0.0045). We then test the resulting hyperplane on systems at lower densities. We find that the classification accuracy is highest for the same density system and that it drops with decreasing density (see Table~\ref{table:highrho}) for the cross-validation accuracies at different densities for this hyperplane.

\begin{table}
\caption{}
\begin{center}
  \begin{tabular}{ | c | c | c | }
 \hline
 Density & Classification Accuracy \\
 \hline
 $\rho = 0.82$  & 85 \%    \\
\hline
$\rho = 0.78$  & 60 \%    \\
\hline
$\rho = 0.72$  & 51 \%    \\
\hline
$\rho = 0.69$  & 50.8 \%    \\
\hline
$\rho = 0.65$  & 50.6 \%    \\
\hline
\end{tabular}
\label{table:highrho}
\end{center}
\end{table}

We then repeat this procedure, assembling the training set from the lowest density system ($\rho=0.65$) at a low temperature (T = 0.00065) where the relaxation time $\tau_\alpha$ is fairly similar to that of the high-density system at $T=0.0045$. The cross-validation accuracy of the resulting hyperplane is shown in Table~\ref{table:lowrho}. In this case, the cross-validation accuracy is significantly lower at the higher densities.

\begin{table}
\caption{}
\begin{center}
  \begin{tabular}{ | c | c | c | }
 \hline
 Density & Classification Accuracy \\
 \hline
 $\rho = 0.82$  & 50 \%    \\
\hline
$\rho = 0.78$  & 50.1 \%    \\
\hline
$\rho = 0.72$  & 51 \%    \\
\hline
$\rho = 0.69$  & 56.5 \%    \\
\hline
$\rho = 0.65$  & 84 \%  \\   
\hline
\end{tabular}
\label{table:lowrho}
\end{center}
\end{table}

Finally, we assemble a training set with half of the particles in the rearranging subset as well as half from the non-rearranging subset taken from the first training set, and the remaining half for each subset taken from the second training set. The resulting hyperplane performs very well in classifying rearranging and non-rearranging particles in a variety of fragile liquids at all the density state points studied (Table~\ref{table:allrho}).

\begin{table}
\caption{}
\begin{center}
  \begin{tabular}{ | c | c | c | }
 \hline
 Density & Classification Accuracy \\
 \hline
 $\rho = 0.82$  & 83 \%    \\
\hline
$\rho = 0.78$  &  82.5\%    \\
\hline
$\rho = 0.72$  &  82.4\%    \\
\hline
$\rho = 0.69$  &  82.1\%    \\
\hline
$\rho = 0.65$  &  82\%    \\
\hline
\end{tabular}
\label{table:allrho}
\end{center}
\end{table}
This hyperplane successfully classifies rearranging/non-rearranging particles based on their local structural descriptors over a wide spectrum of fragile systems. We use this hyperplane throughout our studies to obtain the results reported here. 
The continuous variable softness of particle $i$ ($S_i = w \cdot \mathbf{F_i} - b $) is defined as the signed distance between the hyperplane and the position of local structural environment, as specified by a point in M-dimensional structure-function space.

\begin{figure*}
\centering
\includegraphics[scale=0.18]{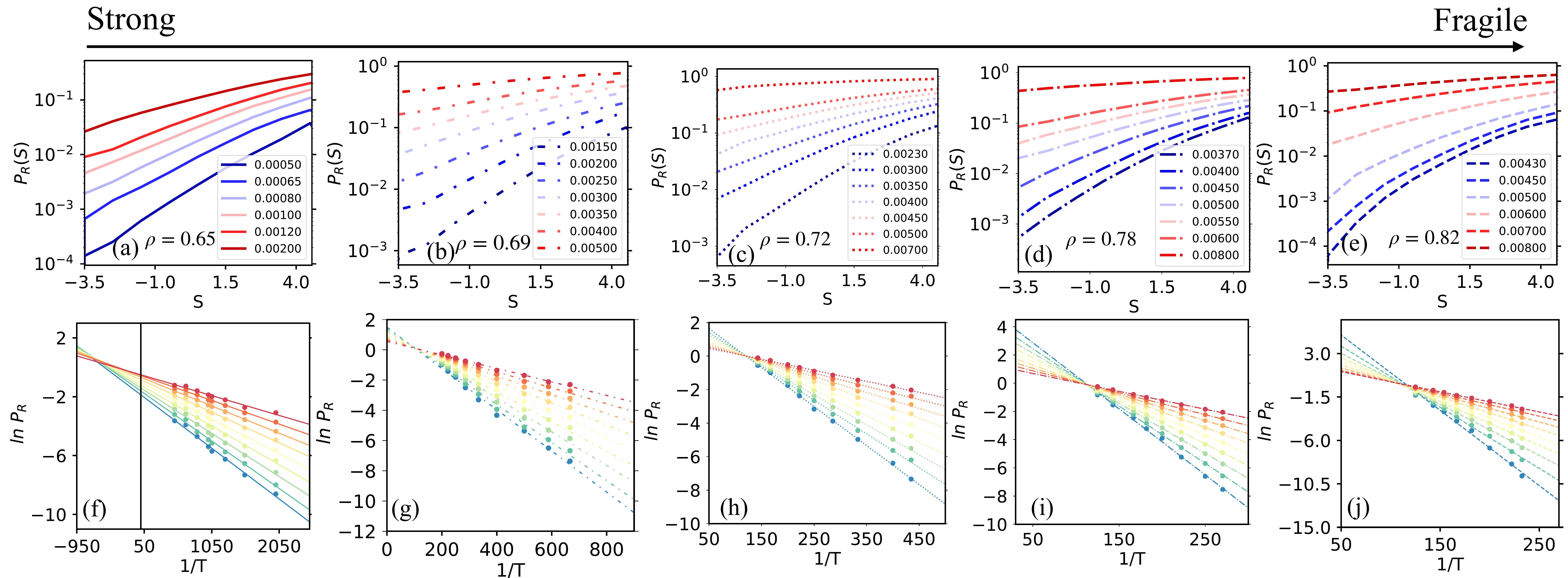}
\caption{The probability of that a particle of softness $S$ will rearrange, $P_R(S)$ vs.~$S$ for systems ranging from strong to  fragile liquids (Fig.1(a) to Fig.1(e)); lines are to guide the eye and temperatures $T$ are indicated in the legend. Arrhenius plots of $P_R(S)$ as a function of $1/T$ for different softnesses ranging from $S \sim -3.5$ (blue) to $S \sim 4.5$ (red) for systems ranging from strong to fragile liquids (Fig.1(f) to Fig.1(j)). Lines are fits to the Arrhenius form (Eq.~\ref{eq:Arrhenius}). Vertical line in Fig.1(f)  indicates the $1/T_0 = 0$ location.}  
  \label{fig:1}
\end{figure*}
\section{\bf Results}
\label{section:4}

\begin{figure*}
\includegraphics[scale=0.13]{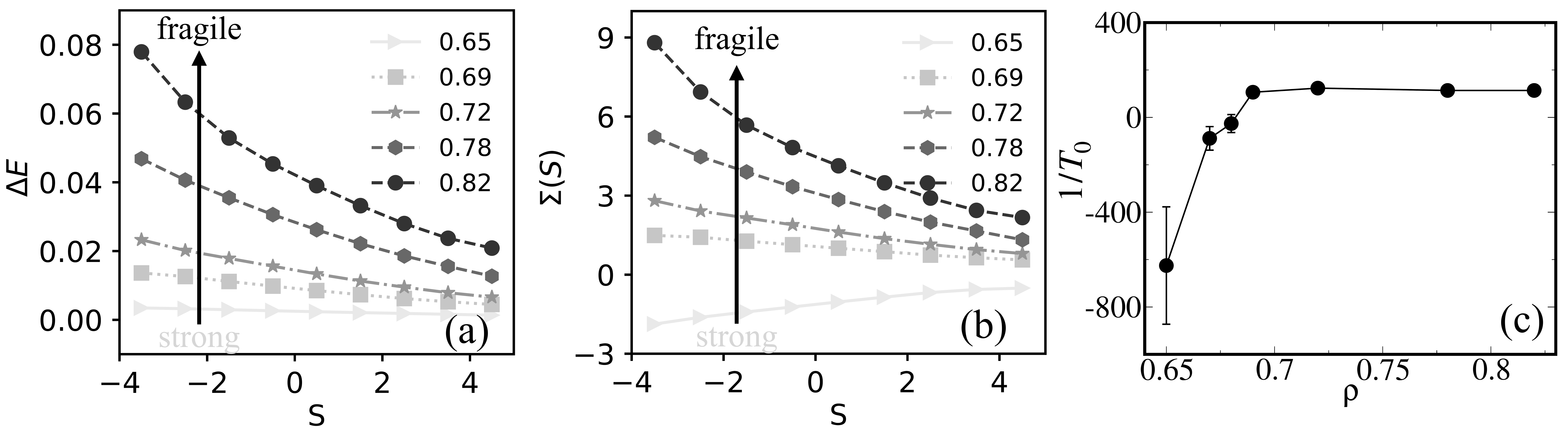}
\caption{(a) Energetic and (b) entropic barriers implied by the Arrhenius behavior of $P_R(S)$ (Eq.~\ref{eq:Arrhenius}) as functions of softness $S$. Densities $\rho$ are indicated in the legend. (c) Inverse onset temperature, $1/T_{0}$ as a function of system density, with fragility increasing with density.
}
  \label{fig:2}
\end{figure*}

It has been shown previously that fragility increases strongly with density in this family of systems~\cite{Berthier_2009,kinetic_fra}. We first recapitulate earlier analyses~\cite{Schoenholz_Nat_Physics} to calculate the probability that a particle is rearranging ($P_R$) as a function of its softness at each density (Fig.~\ref{fig:1} (a) to Fig.~\ref{fig:1} (e)). The probability of rearrangement is the fraction of particles that undergo rearrangement in the interval $\Delta t$ for a given softness $S$. At the lowest temperature studied for each system, $P_R(S)$ increases by several orders of magnitude as softness increases, showing that the softness is highly correlated with the probability of rearranging. Next, 
in Fig.~\ref{fig:1} (f) to Fig.~\ref{fig:1} (j) we plot $P_R(S)$ as a function of $1/T$ for a range of softness values. As in previous studies~\cite{Schoenholz263,Sussman10601,Sharp10943,PhysRevE.101.010602,D0SM01575J} we see that $P_R(S)$ exhibits Arrhenius behavior for a given softness $S$:

\begin{equation}
P_R(S) = \exp \Big(\Sigma (S) -  \frac{\Delta E(S)}{T} \Big)
\label{eq:Arrhenius}
\end{equation}
where the effective configurational entropy barrier $(\Sigma (S))$ and energy barrier $(\Delta E(S))$ terms do not depend on temperature. The values of $(\Delta E(S))$ and $(\Sigma (S))$ are plotted in Fig.~\ref{fig:2} (a) and Fig.~\ref{fig:2} (b). Note that $\Delta E (S)$ and $\Sigma (S)$ become stronger functions of softness with increasing density (increasing fragility). Note also that $\Delta E(S)$ decreases with decreasing density; this is expected since particle overlaps decrease as the density is lowered below the unjamming transition. Furthermore, although $T_g$ decreases at low density, rescaling $\Delta E(S)$ by $T_g$ does not change these trends.

A consistent feature at every density studied is that the Arrhenius curves for $P_R(S)$ at different $S$ intersect at a reasonably well-defined temperature. In previous work, it was shown that this temperature corresponds to the onset temperature, $T_0$, below which liquids begin to show characteristic glassy dynamics, including dynamic heterogeneities, effective barrier heights that increase with decreasing temperature (for fragile systems) and non-exponential relaxation in time~\cite{Schoenholz_Nat_Physics}. The plots for $P_R(S)$ show that $T_0$ corresponds to the temperature at which structure no longer affects dynamics. As $T$ decreases below $T_0$, structure (softness) affects dynamics increasingly strongly.

Figs.~\ref{fig:1}(f)-(j) show that $T_0$ is a function of density, $\rho$. The density dependence is plotted in Fig.~\ref{fig:2}(c). Interestingly, $T_0$ is independent of $\rho$ for sufficiently high $\rho$, corresponding to fragile systems. Below $\rho \approx 0.69$,
however, $1/T_0$ decreases rapidly and even goes negative for the strongest systems studied.

Below $T_0$ (above $1/T_0$) the variation with $S$ of $P_R(S)$ or equivalently the effective energy barrier $\Delta E(S)$, implies that particles of higher softness face lower energy barriers and relax more rapidly, leading to heterogeneity of the relaxation. This heterogeneity is typically characterized using $\chi_4(t)$, the variance of the overlap function ($Q(t)$) 
~\cite{doi:10.1146/annurev.physchem.51.1.99,PhysRevResearch.2.022067} as
\begin{equation}
\chi_4(t) = N \big [\langle Q^2(t) \rangle - \langle Q(t) \rangle ^2 \big].
\label{EQ:1}
\end{equation}
Where the self-overlap correlation function or two-point overlap correlation function is defined as
\begin{equation}
Q(t) = \frac{1}{N} \sum_{i=1}^{N} w\big( |\vec{r}_i(t) - \vec{r}_i(0)| \big).
\label{EQ:2}
\end{equation}
It calculates the amount of overlap between the initial ($t=0$) and subsequent configurations at time $t$. The window function $w(x)$ is defined as $w(x) = 1$, if $x<a$ and $0$ otherwise. The window function $w(x)$ was created to eliminate the decorrelation caused by particles' vibrational motions inside the local cages that their neighbors create. We chose $a=0.3$, which corresponds to the mean square displacement plateau value. The reported results are not sensitive to the precise choice of $a$. 
 
 \vskip +0.05in

\begin{figure}[htp]
\includegraphics[scale=0.245]{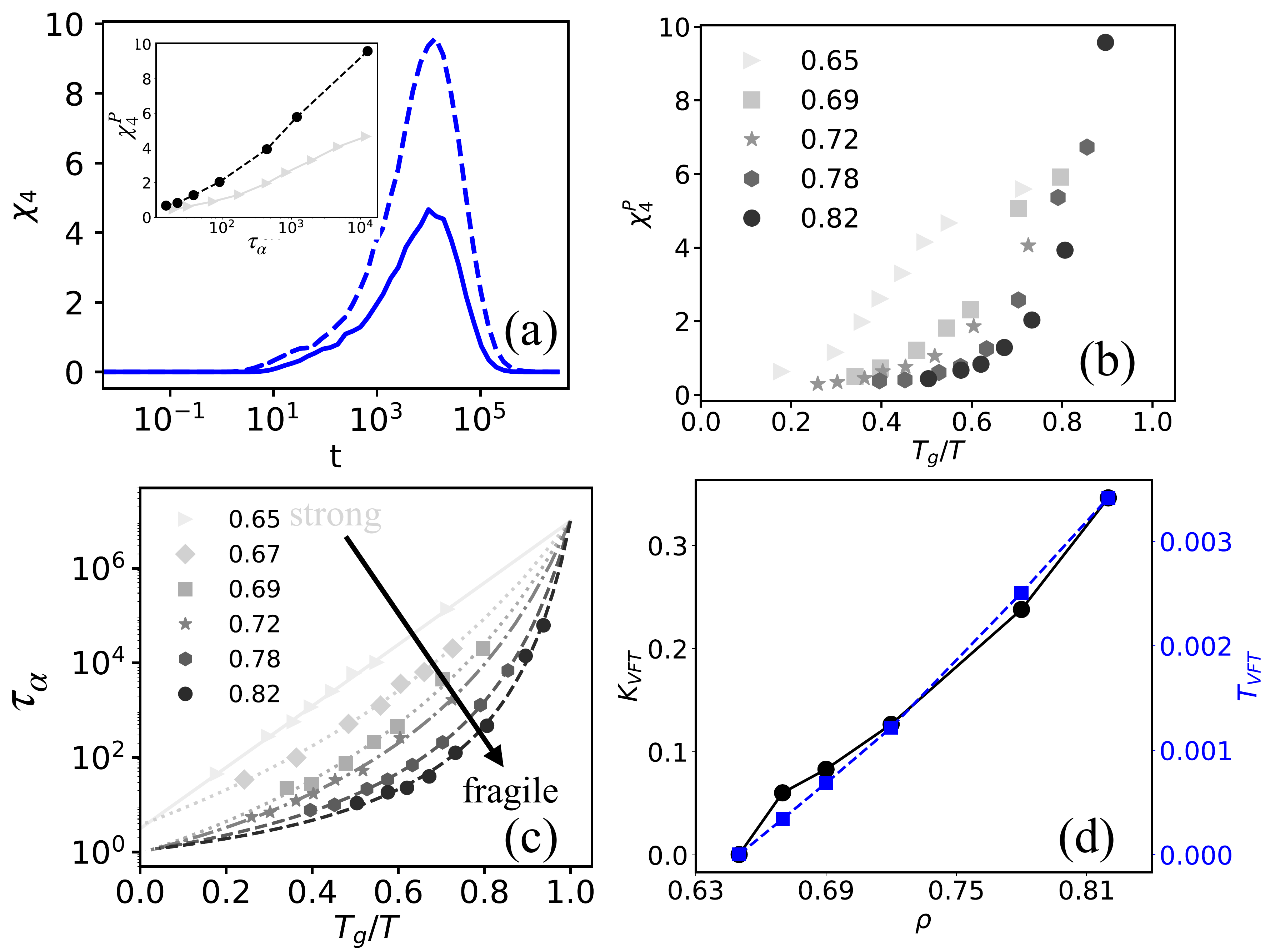}
\caption{(a) The dynamical heterogeneity function $\chi_4(t)$ as a function of time $t$ for strong ($\rho=0.65$, $T = 0.00065$ dotted line) and fragile ($\rho = 0.82$, $T = 0.0045$, solid line) liquids with the same relaxation times. The inset shows the peak value of $\chi_4$, $\chi_4^P$, as a function of the relaxation time $\tau_\alpha$ (time at the peak) for the strongest (solid line) and most fragile (dotted line) liquids studied. (b) The peak value  $\chi_4^P$ is plotted as a function of $T_g/T$, the inverse temperature made dimensionless by the glass transition temperature $T_g$. (c) Angell plot of $\tau_\alpha$ as a function of $T_g/T$. Densities are indicated in the legend. The behavior is Arrhenius at $\rho=0.65$ and becomes increasingly super-Arrhenius with increasing $\rho$. Curves are fits to the Vogel-Fulcher-Tamann form, Eq.~\ref{VFT}. (d) Kinetic fragility  (left y-axis) and $T_{VFT}$ (right y-axis), as defined in Eq.~\ref{VFT}, as a function of density.}
  \label{fig:3}
\end{figure}

Fig.~\ref{fig:3}(a) shows representative examples of $\chi_4(t)$ for the highest- and lowest-density systems studied, showing that there is a peak in this quantity. The peak position, $\tau_\alpha$, is a measure of the relaxation time, while peak magnitude, $\chi_4^P$ quantifies the degree of dynamical heterogeneity. 

The temperature dependences of $\chi_4^P$ and $\tau_\alpha$ are shown in Fig.~\ref{fig:3}(b)-(c), where we have rescaled $T$ by the corresponding glass transition temperature ($T_g$) for the system at each density. Here we define the glass transition temperature, $T_g$ as the temperature at which $\tau_\alpha = 10^7$. Values of $T_g$ are listed in Table \ref{table:Tg}.
\begin{table}
\caption{}
 \begin{center}
  \begin{tabular}{ | c | c | c | }
 \hline
 Density & $T_g$ \\
\hline
$\rho = 0.65$  &  0.000357   \\
\hline
$\rho = 0.69$  &  0.001195   \\
\hline
$\rho = 0.72$  &  0.001812    \\
\hline
$\rho = 0.78$  &  0.003163   \\
\hline
 $\rho = 0.82$  & 0.004031  \\
 \hline
\end{tabular}
\label{table:Tg}
\end{center}
\end{table}

Fig.~\ref{fig:3}(c) shows that fragility increases with density, as shown earlier in Ref.~\cite{Berthier_2009,kinetic_fra}. We fit the relaxation time $\tau_\alpha(T)$ using the Vogel-Fulcher-Tamann (VFT) formula \cite{10004192038,https://doi.org/10.1111/j.1151-2916.1925.tb16731.x,https://doi.org/10.1002/zaac.19261560121} 
\begin{equation}
\tau_\alpha(T)=\tau_0 \exp\left [\frac{1}{K_{VFT} \ (T/T_{VFT}-1)}\right],
\label{VFT}
\end{equation}
where $\tau_0$ is the relaxation time at the onset temperature $T_0$. Here $T_{VFT}$ is the apparent divergence temperature for relaxation time and $K_{VFT}$ determines the degree of kinetic fragility, known as ``fragility index''. In Fig.~\ref{fig:3}(c) the curves are the VFT fits. Fig.~\ref{fig:3}(d) shows the kinetic fragility ($K_{VFT}$) (left axis, black points) and $T_{VFT}$ (right axis, blue points) as a function of density ($\rho$). We see that kinetic fragility ($K_{VFT}$) and $T_{VFT}$ provide essentially the same information and increase with increasing fragility.

Returning to Fig.~\ref{fig:3}(a), we note that it shows $\chi_4(t)$ for the strongest (solid line) and most fragile (dotted line) systems (the lowest- and highest-density systems), with temperatures chosen for each so that the two systems have the same value of $\tau_\alpha$. In the inset we plot $(\chi_4^P)$  as a function of $\tau_\alpha$ for the two systems. Evidently the more fragile system has a higher value of $\chi_4^P$ than the stronger system at the same relaxation time, as expected from the stronger dependence of $\Delta E$ and $\Sigma$ on $S$ for the more fragile system. 

\begin{figure}[htp]
\includegraphics[scale=0.126]{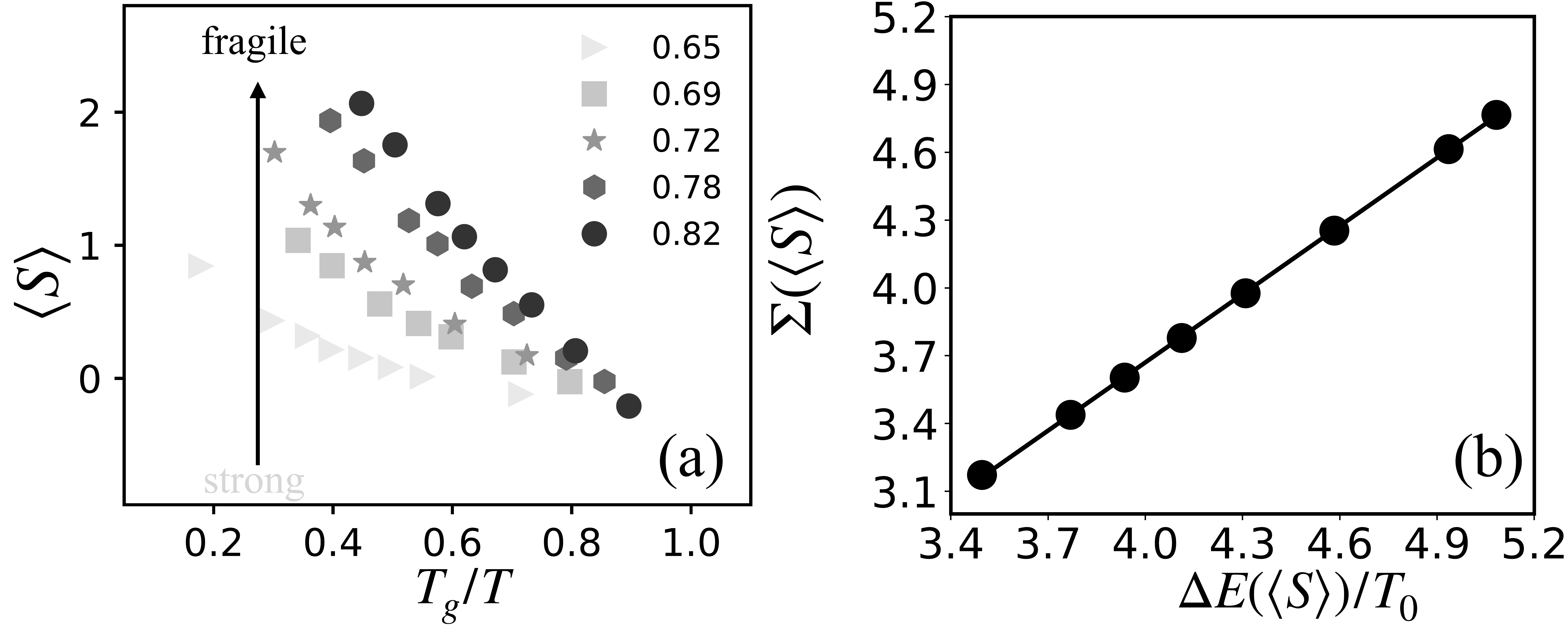}
\caption{(a) Average softness $\langle S \rangle$ as a function of $T_g/T$ for systems at densities indicated in the legend.  The temperature dependence of $\langle S \rangle$ increases with increasing fragility (density). 
(b) The relation between $\Sigma(\langle S \rangle)$ and $\Delta E(\langle S \rangle)/T_0$ (described after Eq.~\ref{eq:sigmavsdeltae}) for $\rho=0.82$.}
  \label{fig:4}
\end{figure}

For fragile liquids the relaxation time increases more rapidly with decreasing temperature $T$ than for strong liquids. Because softness corresponds to an energy barrier, one might expect that the mean softness, $\langle S \rangle$ of a fragile system is a stronger function of temperature than that of a strong system. In Fig.~\ref{fig:4} (a), we plot $\langle S \rangle$ as a function of $T_g/T$. For fragile liquids, $\langle S \rangle$ decreases at a significantly faster rate with decreasing temperature than for strong liquids, as expected. Note that the definition of softness is identical for all systems studied, so this result indicates that structure changes more rapidly with temperature, leading to a more rapid change of relaxation time in fragile liquids than in strong ones.

To better understand the connection between structure, energy barrier and dynamics, we study the free energy barrier:
\begin{equation}
    \Delta F(S) = \Delta E(S) - T \Sigma (S)
    \label{eq:free_energy}
\end{equation}

\begin{figure}[htp]
\includegraphics[scale=0.13]{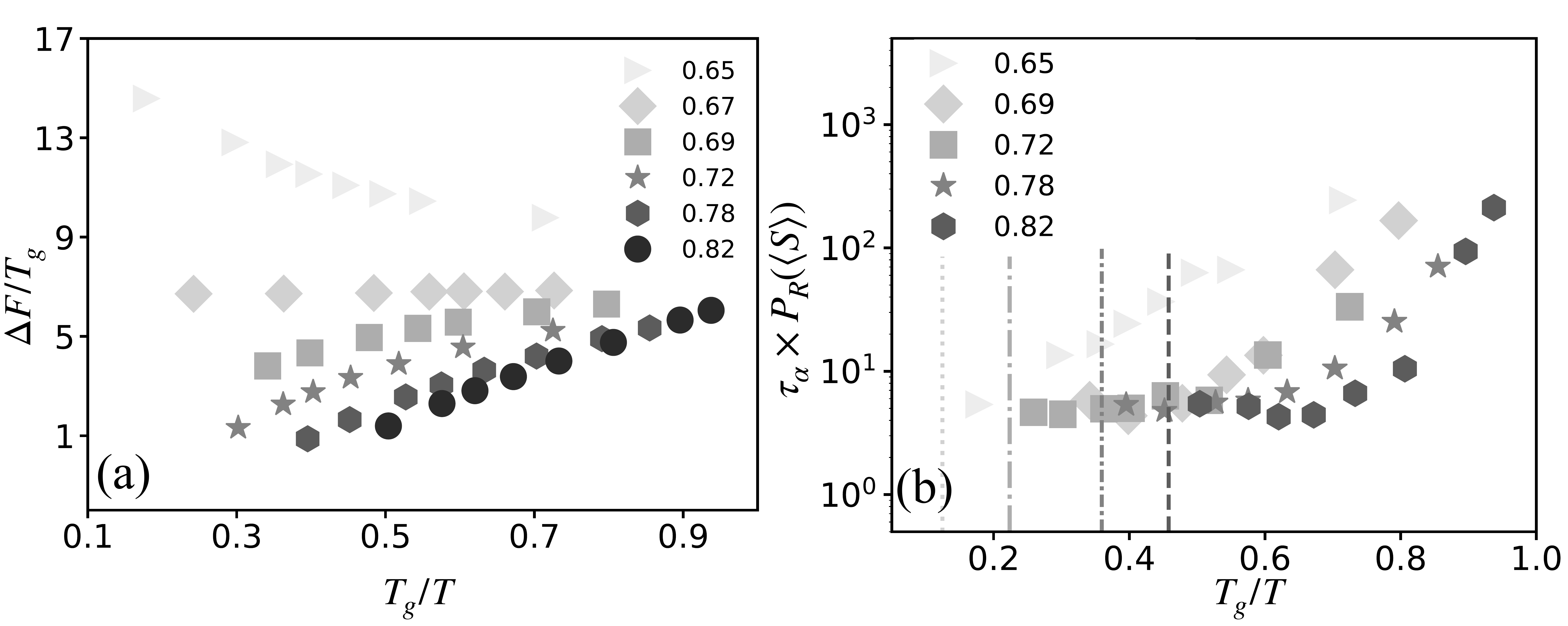}
\caption{(a) Data for the dimensionless free energy barrier $(\Delta F(\langle S \rangle)/T_g)$ for systems of different densities against inverse temperature ($T_g/T$). (b) Product of $\tau_\alpha$ and $P_R(\langle S \rangle )$ against inverse temperature ($T_g/T$). Vertical lines represent the re-scaled onset temperature ($T_g/T_0$) for systems at densities $\rho=0.69$ to $\rho=0.82$.}
  \label{fig:5}
\end{figure}

Note that at the onset temperature, $T_0$, $P_R(S)$ is independent of softness so from Eq.~\ref{eq:Arrhenius} we have
\begin{equation}
\Sigma(S) = \Delta E(S)/T_0 + C^\prime. 
    \label{eq:sigmavsdeltae}
\end{equation}
where $C^\prime$ is a constant. In particular, at any temperature $T$, Eq.~\ref{eq:sigmavsdeltae} implies that $\Sigma(\langle S \rangle)= C^\prime + \Delta E(\langle S \rangle)/T_0$. This relation is verified in Fig.\ref{fig:4}(b) for the most fragile system, $\rho=0.82$; we find excellent agreement with $C^\prime=-0.35$.
Combining this with Eq.~\ref{eq:free_energy} we see that 
\begin{equation}
    \Delta F(S)=(1-T/T_0) \Delta E(S) - T C^\prime.
      \label{eq:free_energy_new}
\end{equation}
If we were to assume that $\tau \propto 1/P_R(\langle S \rangle)$ as suggested by earlier work~\cite{Schoenholz263},  this would imply 
$\tau \propto \exp({\Delta F(\langle S \rangle)}/T)$. We check the relation $\tau \propto 1/P_R(\langle S \rangle)$ by plotting the product $\tau_\alpha \times P_R(\langle S \rangle)$ as a function of inverse temperature in Fig.~\ref{fig:5} (b). We see that the product is not constant with temperature, implying that deviations from the simple relation $\tau \propto 1/P_R(\langle S \rangle)$. These deviations are pushed to lower temperatures with increasing fragility (density), partly due to the decrease in $T_0$, indicated by the vertical lines in Fig.~\ref{fig:5}(b). Thus, deviations are compressed into a narrower range, leading to a stronger $T$-dependence of the relaxation time, with increasing fragility. In all cases, however, such deviations are observed, showing that the dynamics are not simple, likely due to spatial and temporal correlations such as those that lead to kinetic heterogeneities (\textit{e.g.}~facilitation), which are stronger for more fragile systems (Fig.~\ref{fig:3}(a)). 

\section{\bf Discussion}
\label{section:5}

In summary, our results show that structure makes a strong contribution to the wide variation in fragility observed in glassy liquids. Our analysis suggests that three factors control fragility. Compared to strong liquids, in fragile ones (1) local structure ($\langle S \rangle$) decreases more strongly with decreasing temperature, (2) the free energy barrier $\Delta F(\langle S \rangle)$ rises more rapidly with decreasing $\langle S \rangle$, and (3) the upwards deviation of $\tau$ from $1/P_R(\langle S \rangle)$ is compressed into a smaller temperature range relative to $T_g$.  All of these factors contribute to the greater sensitivity of the relaxation time $\tau_\alpha$ to temperature in fragile liquids. 

Our interpretation of $\Sigma(S)$ and $\Delta E(S)$ as entropic and energetic barriers stems from the Arrhenius behavior of $P_R(S)$. Direct confirmation that these quantities can be interpreted as barriers is difficult because calculations of energy barriers are extremely costly in a system with $O(e^{O(N)})$ minima once the number of particles, $N$, is of reasonable size. Direct calculations of entropic barriers are even more challenging but could be worthwhile in small systems to compare against our results.

The result of Ref.~\cite{Schoenholz_Nat_Physics} that $T_0$ marks the onset of the dependence of $P_R(S)$ on softness, or local structure, is surprisingly powerful. It allows direct extraction of the entropic barrier (Eq.~\ref{eq:sigmavsdeltae}) and free energy barrier (Eq.~\ref{eq:free_energy_new}) to rearrangement that only include an temperature independent constant, given the energy barrier. Note that $T_0$ marks a crossover, not a transition, in low dimensions, so it is not sharply defined. Nonetheless, our results show that it is sufficiently well defined to be potentially useful as a way of extracting barriers.

Remarkably, we find that $T_0$ is actually negative for the strongest glassformers. Eq.~\ref{eq:sigmavsdeltae} shows that when $T_0$ is positive, a decreasing energy barrier with increasing $S$ implies an increasing entropic barrier with $S$. Thus, softer particles have lower energetic barriers and higher entropic barriers. However, when $T_0<0$, the opposite is true--softer particles have lower energy barriers and lower entropic barriers as well.
Note that for the strongest system at $\rho=0.65$, the energy barrier is essentially independent of softness (Fig.~\ref{fig:2}(a) ) and $T_0<0$ with a very small magnitude (Fig.~\ref{fig:2}(c)). As a result, the heterogeneity in particle mobilities comes from variation of the entropic barriers with $S$ (Fig.~\ref{fig:2}(b)), not of the energy barriers. In other words, at high densities/fragilities, heterogeneous mobility stems from the softness dependence of the energy barriers while at low densities/fragilities, it stems from the softness dependence of the entropic barriers. It is notable that a single hyperplane is predictive at all these densities and captures this change of behavior.

\subsection*{Conflicts of interest}
There are no conflicts of interest to declare.

\subsection*{Acknowledgements}
This research was supported by NSF through the University of Pennsylvania Materials Research Science and Engineering Center (MRSEC) (DMR-1720530) (IT) and through the Extreme Science and Engineering Discovery Environment (XSEDE) \cite{xsede} via TG-BIO200091. Additional support was provided by the Simons Foundation through the Simons Collaboration on ``Cracking the glass problem'' (Award 454945 to SAR and AJL) and through Investigator Award 327939 to AJL.

\subsection*{Data Availability Statement}
Raw data used to make plots and other data will be published at the following url [???] after the revision based on reviewer comments.
 \appendix
\section{Results for the large particles}
\label{Appendix_sce_1}
 \vskip +0.05in
\begin{figure*}
\centering
\includegraphics[scale=0.18]{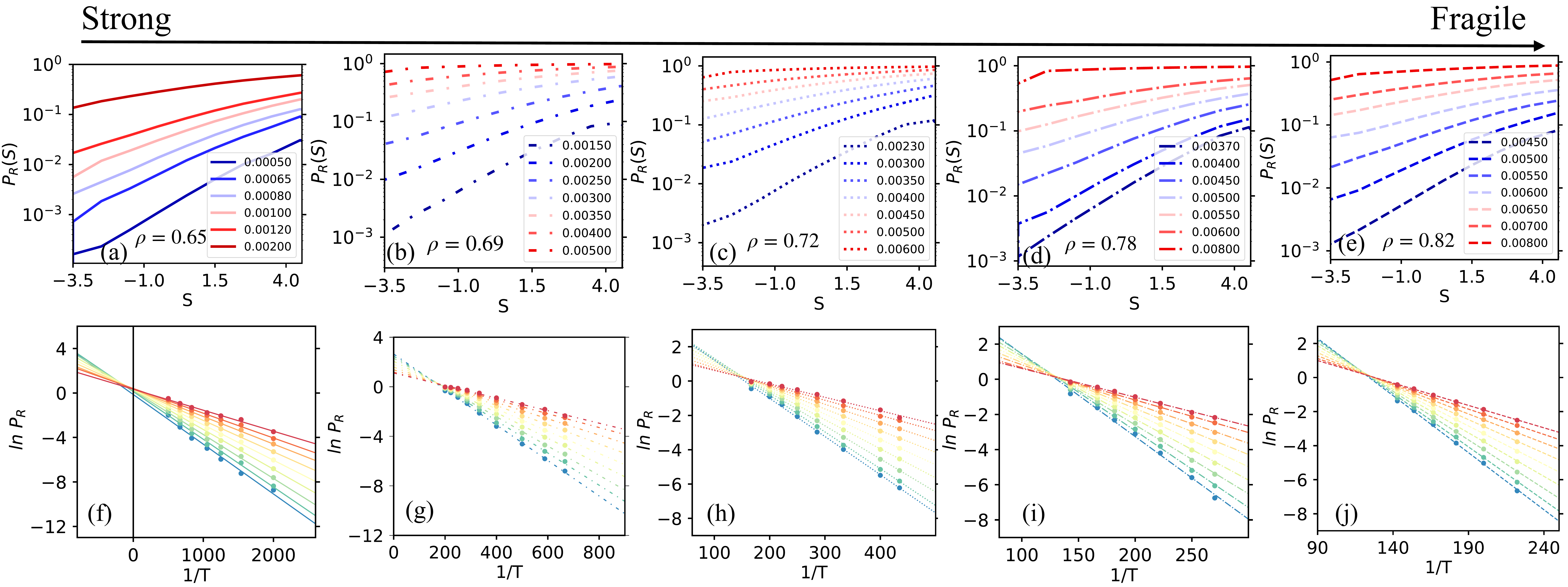}
\caption{The probability of that a particle of softness $S$ will rearrange, $P_R(S)$ vs.~$S$ for systems ranging from strong to  fragile liquids (Fig.6(a) to Fig.6(e)); lines are to guide the eye and temperatures $T$ are indicated in the legend. Arrhenius plots of $P_R(S)$ as a function of $1/T$ for different softnesses ranging from $S \sim -3.5$ (blue) to $S \sim 4.5$ (red) for systems ranging from strong to fragile liquids (Fig.6(f) to Fig.6(j)). Lines are fits to the Arrhenius form (Eq.~\ref{eq:Arrhenius}).} 
  \label{fig:Appendix_1}
\end{figure*}
\vskip +0.05in
\begin{figure}[htp]
\includegraphics[scale=0.22]{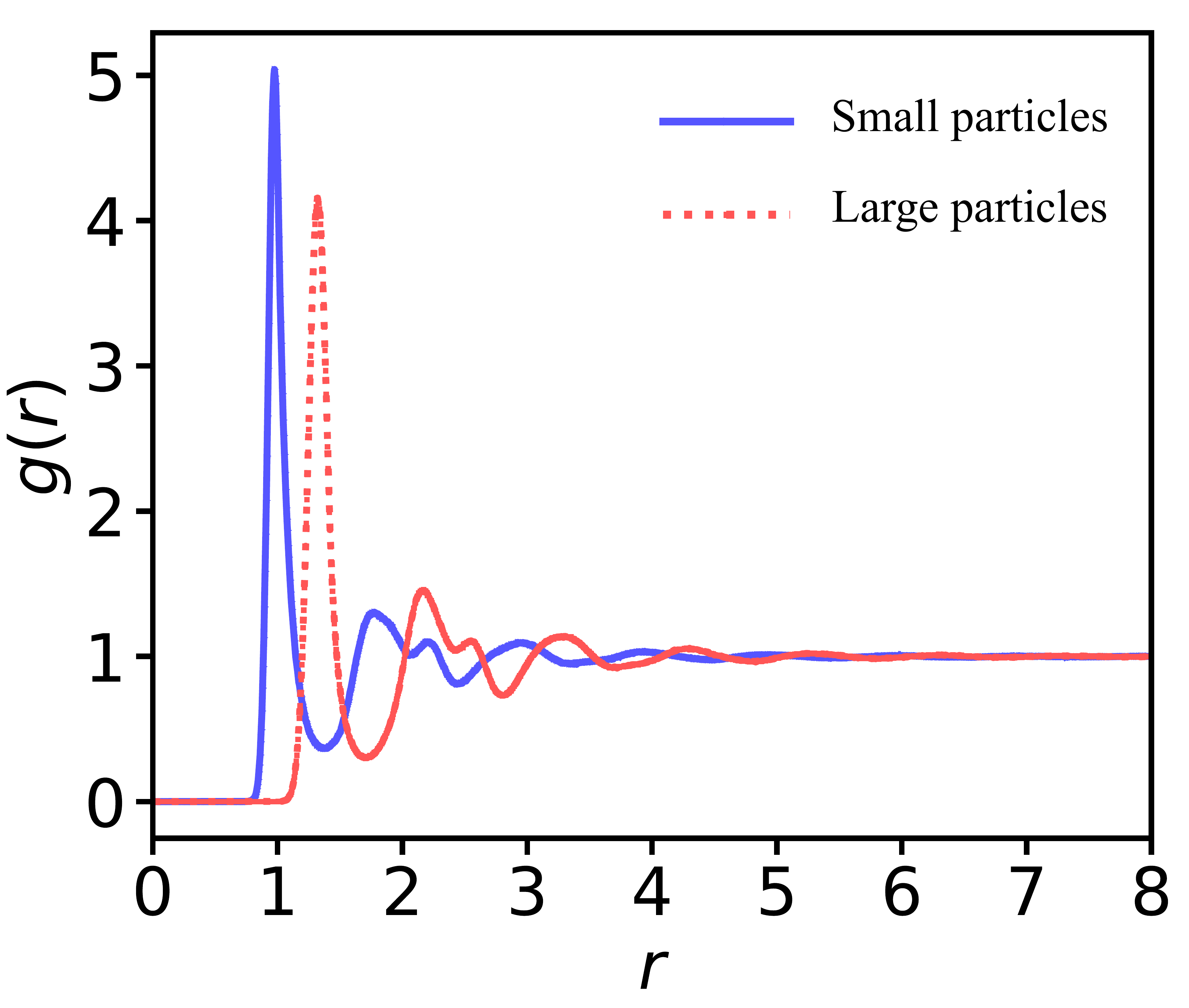}
\caption{Pair correlation functions for large and small particles at density $\rho = 0.82$ and $T = 0.0045$.}
  \label{fig:Appendix_2}
\end{figure}

\begin{figure}
\includegraphics[scale=0.24]{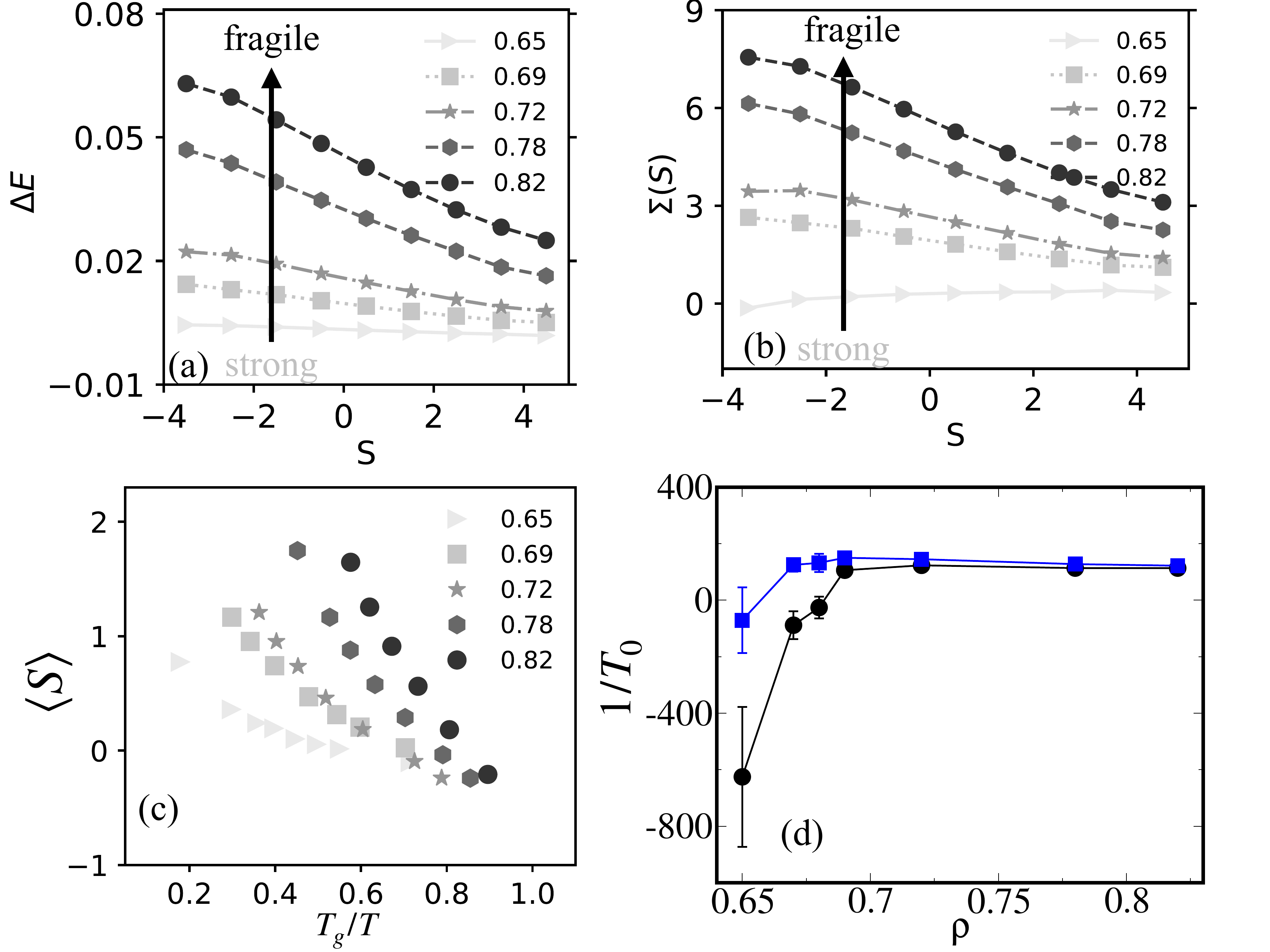}
\caption{(a) The values of energetic barriers ($\Delta E$) (defined in equation.~\ref{eq:Arrhenius}) as a function of softness S. (b) The values of entropic barriers ($\Sigma$) (defined in equation.~\ref{eq:Arrhenius}) as a function of softness S. Densities $\rho$ are indicated in the legend. (c) The average softness ($\langle S \rangle$) as a function of rescaled temperature (rescaled by corresponding glass transition temperature ($T_g$)). (d) Inverse onset temperature, $1/T_0$ as a function of system density, with fragility increasing with density. black circle symbols for small particles and blue square symbols for large particles.} 
  \label{fig:Appendix_3}
\end{figure}

\begin{figure}[htp]
\includegraphics[scale=0.20]{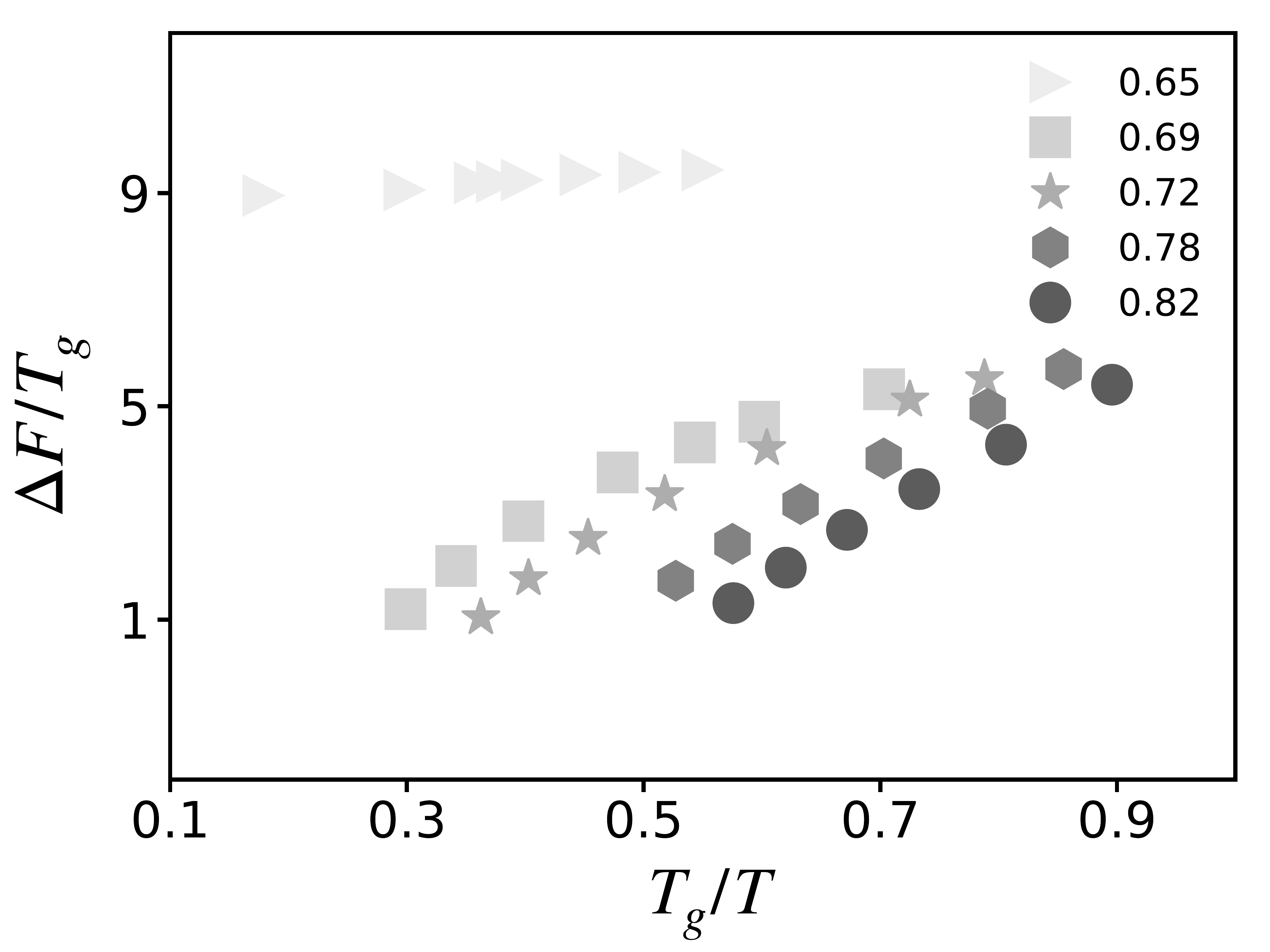}
\caption{Data for the dimensionless free energy barrier $(\Delta F(\langle S \rangle)/T_g)$ for systems of different densities against inverse temperature ($T_g/T$).}
  \label{fig:Appendix_4}
\end{figure}
The main text focuses on the small particles in our binary system. To test the robustness of our results, we performed all of the analyses for the large particles as well. The pair correlation in Fig.~\ref{fig:Appendix_2} shows that large and small particles have different local densities. We construct the structural descriptors, training set, and hyperplane for large particles using the same method as before. The reference particle ($\kappa$) is now a large particle in Equations \ref{density} and \ref{angular}. We obtain a classification accuracy of roughly 84 \% for large particles across all different fragile liquids. The results and plots for large particles are shown below.

In Fig.~\ref{fig:Appendix_1} (a) to  Fig.~\ref{fig:Appendix_1} (e) we plot $P_R(S)$ as a function of particle softness S for large particles. The results are qualitatively the same as for small particles. In Fig.~\ref{fig:Appendix_1} (f) - (j), we show Arrhenius plots of $P_R(S)$ confirming Arrhenius temperature dependence for a given softness S over all systems studied. Again, our results are qualitatively identical for large and small particles.

We further extract $\Delta E(S)$ and $\Sigma (S)$ using Eq.~\ref{eq:Arrhenius}. We plot the values of $\Delta E(S)$ and $\Sigma (S)$  in Fig.~\ref{fig:Appendix_3} (a) and Fig.~\ref{fig:Appendix_3} (b). Both these become a stronger function of softness as fragility increases, just as shown for small particles in Fig.~\ref{fig:2}(a-b).  In Fig.~\ref{fig:Appendix_3} (c) we plot average softness ($\langle S \rangle$) as a function of $T_g/T$. For fragile liquids, the average softness drops considerably faster with decreasing temperature than for strong liquids, as in Fig.~\ref{fig:3}(a). The density dependence of onset temperature ($T_0$) for large (blue square symbols) and small particles (black red symbols) is plotted in (Fig.~\ref{fig:Appendix_3}(d). Interestingly, onset temperatures ($T_0$) for large and small particles are very similar and independent of $\rho$ for high 
$\rho$, however for strongest systems, $T_0$ starts to separate for large and small particles and goes negative. 
In Fig.~\ref{fig:Appendix_4} we plot $\Delta F(\langle S \rangle)/T_g$ as a function of $T_g/T$. Similarly, $\Delta F(\langle S \rangle)/T_g$ is a stronger function of temperature for fragile than for strong liquids (Fig.~\ref{fig:Appendix_3}(d), as it is for small particles (Fig.~\ref{fig:5}(a)). 

\balance
\clearpage
\vskip +0.05in
\bibliography{Paper_fragility} 
\bibliographystyle{ieeetr}
\end{document}